\documentclass[%
prapplied,
twocolumn,
amsmath,
amssymb,
reprint,
superscriptaddress,
footinbib,
 amsmath,amssymb,
 aps,
longbibliography
]{revtex4-2}
\usepackage[colorlinks=true,citecolor=blue,linkcolor=black]{hyperref}
\usepackage{romannum}
\usepackage{lipsum}
\usepackage{float}
\usepackage{graphicx}
\usepackage[utf8]{inputenc}
\usepackage[english]{babel}
\usepackage{amsmath,amsfonts,amssymb}
\usepackage[T1]{fontenc}
\usepackage{url}
\usepackage{changes}
\usepackage{amsmath}
\usepackage{amsfonts}
\usepackage{amssymb}
\usepackage[version=3]{mhchem}
\usepackage{changes}
\definechangesauthor[color=red]{TJK}
\usepackage{changes}
\usepackage{soul}

\begin{document}

\title{Photonic electrometry using a piezoelectric-Pockels microresonator}

\author{Suwan Sun}
\affiliation{Key Laboratory of Specialty Fiber Optics and Optical Access Networks, Joint International Research Laboratory of Specialty Fiber Optics and Advanced Communication, Shanghai University, Shanghai 200444, China}
\affiliation{Institute for Photonics and Advanced Sensing (IPAS) and School of Physics, Chemistry and Earth Sciences, University of Adelaide, Adelaide SA 5005, Australia}

\author{Hairun Guo}
\affiliation{Key Laboratory of Specialty Fiber Optics and Optical Access Networks, Joint International Research Laboratory of Specialty Fiber Optics and Advanced Communication, Shanghai University, Shanghai 200444, China}

\author{Andre N. Luiten}
\affiliation{Institute for Photonics and Advanced Sensing (IPAS) and School of Physics, Chemistry and Earth Sciences, University of Adelaide, Adelaide SA 5005, Australia}
\affiliation{ARC Centre of Excellence in Optical Microcombs for Breakthrough Science (COMBS), University of Adelaide, Adelaide, South Australia, 5005, Australia}

\author{Wenle Weng}
\email[]{wenle.weng@adelaide.edu.au}
\affiliation{Institute for Photonics and Advanced Sensing (IPAS) and School of Physics, Chemistry and Earth Sciences, University of Adelaide, Adelaide SA 5005, Australia}
\affiliation{ARC Centre of Excellence in Optical Microcombs for Breakthrough Science (COMBS), University of Adelaide, Adelaide, South Australia, 5005, Australia}

\begin{abstract}
Facilitated by low-noise laser frequency locking, optical microresonators with the Pockels effect have shown unprecedented high resolutions in sensing electrical field. However, the requirement for tunable and low-noise laser sources considerably increases the cost and the size of the system, thereby limiting the industrial applicability of the microresonator-based technology. Here, we explore the possibility of using a low-cost fixed-frequency semiconductor laser as the pump laser to perform radiofrequency electrometry. A resonant mode in a lithium niobate microresonator is frequency-locked to the laser using the electrooptic effect. This same effect also underlies the radiofrequency electric-field sensing mechanism. Our experimental results show that the electrometry resolution can be maintained at signal frequencies beyond the optical resonance bandwidth and that the signal-to-noise ratio does not change with varied coupling conditions as long as the laser frequency noise is the dominant noise source of the system. In addition, narrowband electrooptic sensitivity enhancement is observed at frequencies of the microresonator's piezoelectric resonances, resulting in a resolution enhancement factor of approximately 3 at signal frequencies around 4 MHz. Our work advances the photonic resonant electrometry technology by studying the bandwidth limitation, and opens the road to the employment of low-cost lasers in high-resolution sensing applications.
\end{abstract}

\maketitle
%%%%%%%%%%%%%%%%%%%%%%%%%%%%%%
%\section{Introduction}
%\textit{Introduction.}---
%%%%%%%%%%%%%%%%%%%%%%%%%%%%%%
\section{Introduction}

Optical microresonators have become an attractive platform for sensing applications \cite{Yu:2021aa,ward2011WGM} due to their high quality-factors ($Q$s), small footprints, and highly localized optical modes. These attributes give rise to strong light-matter interactions and high sensitivities to environmental perturbations, making microresonators ideal candidates for high-performance sensor development \cite{Hsu:2007aa,zhu2009chip,weng2015refractometry,Chen:12,yu2022single,PhysRevApplied.5.044007,Yang:16,Weng:18,Heylman:2016aa,Cao:2024aa}. Among different sensor types, optical and photonic electric-field (E-field) sensors have gained research interest due to their crucial application values in both fundamental scientific research \cite{Sedlacek:2012aa,Zhang:2020aa} and industrial usages \cite{han2021micro,Ma:2024aa,wang2019transient} such as material characterization, advanced manufacturing, medical diagnosis, and environmental monitoring. Compared to conventional electrostatic-force-based sensors \cite{Kainz:2018aa,han2022trampoline}, optical and photonic E-field sensors exhibit high sensitivities, low absorptions, high bandwidth, and reduced vulnerabilities to interferences \cite{Zhu:2015aa,Michl:2019aa,yang2019intense,liu2022highly}. Photonic materials with a low optical absorption and the electrooptic (Pockels) effect are ideal for the development of photonic E-field sensors. Among various materials such as barium titanate (BaTiO$_3$) \cite{10.1063/1.1498151,7395309}, lithium tantalate (LiTaO$_3$) \cite{Casson:04}, aluminium nitride (AlN) \cite{7065215}, and silicon carbide (SiC) \cite{Wang:23}, lithium niobate (LN) is likely to be the most attractive due to its high electrooptic coefficients \cite{Yonekura_2008}, material stability, and fabrication maturity. Recently, thin-film LN photonic microresonators have been used for radiofrequency (rf) E-field sensing \cite{Ma:2024aa}. Using the Pound-Drever-Hall (PDH) technique \cite{black2001introduction,drever1983laser} to precisely read out the optical resonance frequency variations translated from the external E-field strength, an impressive resolution of 5.2 $\mu \text{V}/(\text{m}\sqrt{\text{Hz}})$ has been achieved with strong signal amplification enabled by the dipole antenna. However, in order to produce the low-noise PDH signal that is necessary for achieving a high sensing resolution, a high-performance tabletop laser with a low frequency noise is required, which not only considerably increases the cost but also to a great extent cancels the compactness and portability yielded by the microresonator technology.

In this work, we use a crystalline LN microresonator driven by a fixed-frequency semiconductor laser for antenna-free rf electrometry. As the low-frequency end of the sensing bandwidth in our experiment is below 10 kHz, the removal of the antenna not only avoids the distortion of the E-field profile due to the antenna resonance and radiation but also reduces the complexity of the device. To enable sensing, the resonance frequency of a microresonator mode is locked to the laser using electrooptic feedback-control directly applied to the microresonator. At frequencies beyond the locking bandwidth, the E-field strength is measured by monitoring the in-loop PDH error signal. We show that, with the laser frequency noise being the dominant noise source of the PDH system, the same sensing resolution can be maintained at frequencies beyond the optical resonance bandwidth. Additionally, at the frequencies of the piezoelectric resonances of the LN crystal, the sensing resolution is enhanced due to the increase of the effective electrooptic response \cite{takeda2012electro,weng2024low,garzarella2005piezo,LI2018523}. Supported by theoretical analysis, our proof-of-principle experiments provide a widely applicable strategy to the design and development of cost-effective resonant photonic sensors based on lasers with limited tunability and relatively high frequency noises.

%%%%%%%%%%%%%%%%%%%%%%%%%%%%%%%%%%%%%%%%%%%%%%%%%%%%%%%%%%%%%%%%%%%%%%%%%%%%%%%%%%%%%%%%%%%%%%%%%%%%%%%%%%%%%%%%%%%%%%%%%%%%%%%%%%%%%%%%%%%%%%%%%%%%%%%%%%%%%%%%%%%%%%%%%%%%%%%%%%%%%%%%%%%%%%%%%%%%%%%%%%%%%%%%%%%%%%%%%%%%%%%%%%%%%%%%%%%%%%%%

\section{Experiments}

\subsection{Characterization of the LN microresonator}

%%%%%%%%%%%%%%%%%%%%%%%%%%%%%%%%%%%%%%%%%%%%
\begin{figure*}[hbt]
\centering
\includegraphics[width=1.6\columnwidth]{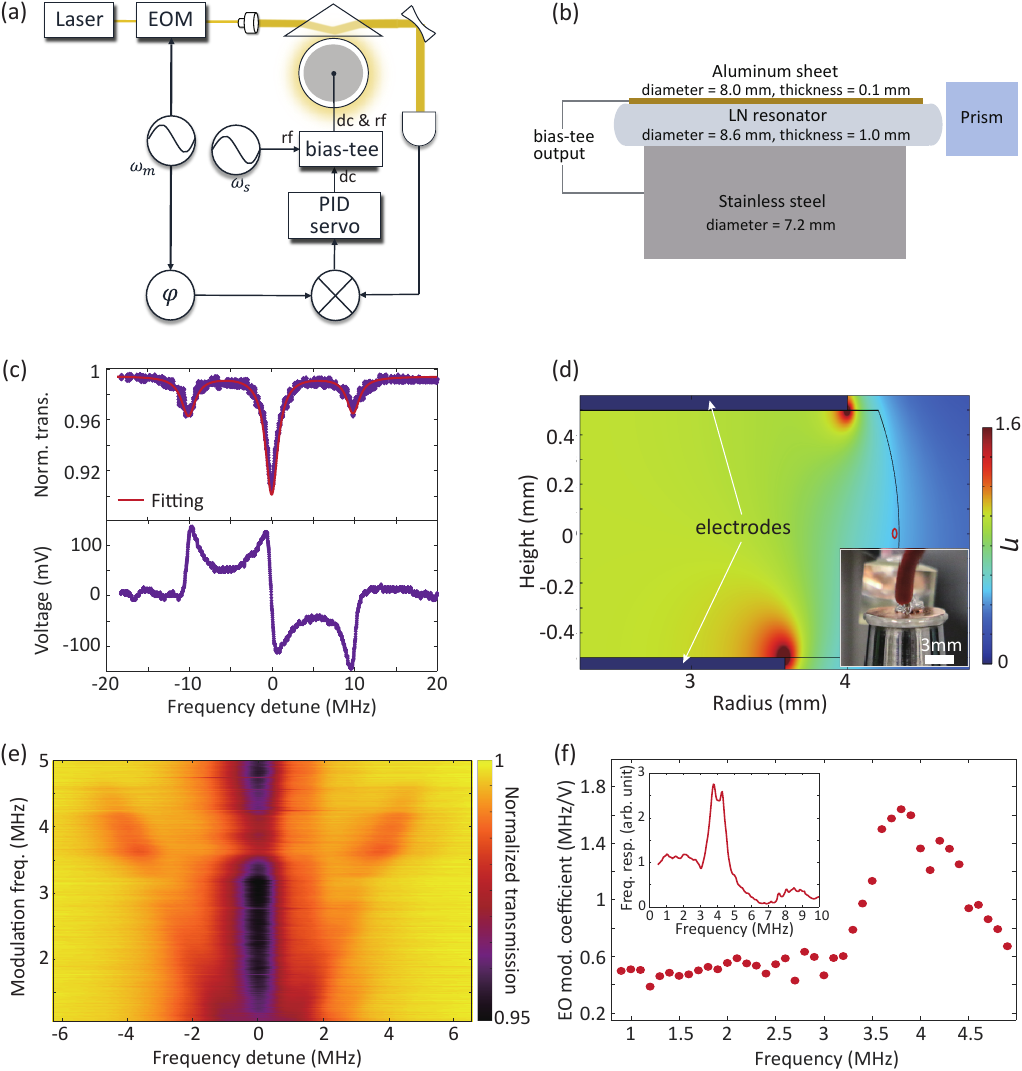} 
\caption{
(a) Main diagram of the experimental setup. The angular frequency $\omega_\mathrm{m}$ is the laser-phase-modulation frequency. The frequency of the rf signal applied to the microresonator is denoted as $\omega_\mathrm{s}$.
(b) The detailed configuration of the whispering-gallery-mode resonator. The dimensions of the microresonator and the electrodes are shown.
(c) The transmission spectrum of a resonance with a $Q$ of approximately $1\times10^8$ and the corresponding PDH error signal.
(d) Simulated electrode efficiency. The red-circled oval indicates the location of the optical mode. The inset shows the photo of the microresonator.
(e) Laser-swept transmission spectrogram of an electrooptically modulated resonance, showing the sidebands generated by the frequency modulation of the resonance. The magnitude of the sidebands becomes abruptly strong around the modulation frequency of 4 MHz due to piezoelectric resonances of the microresonator.
(f) Electrooptic modulation coefficients derived with the spectrogram in (e). The inset is the electrooptic modulation response profile of the microresonator measured by a VNA.
}\label{fig1}
\end{figure*}
%%%%%%%%%%%%%%%%%%%%%%%%%%%%%%%%%%%%%%%%%

The experimental setup is shown in Fig. \ref{fig1} (a). A semiconductor laser (Photonx PHX-C-F-B-D34) with a fixed wavelength at 1549.6 nm is phase-modulated at 10 MHz by an electrooptic modulator (EOM) and then coupled into a LN whispering-gallery-mode microresonator using a rutile prism. The microresonator with a diameter of 8.6 mm and a thickness of 1 mm is fabricated by the traditional method of surface polishing \cite{Weng:24} with a workshop lathe. Sandpapers are first used to shape the microresonator from a Z-cut LN disk, and then fine polishing with diamond slurry is applied to achieve a smooth microresonator surface. After the fine polishing, the surface is cleaned with water and acetone. A copper sheet and a stainless-steel post are attached to the top and the bottom of the microresonator respectively so external electrical field can be applied on the microresonator with these electrodes (see Fig. \ref{fig1} (b)). For the PDH error signal generation, the transmission light is registered by a high-bandwidth photodetector and the rf output of the photodetector is demodulated by the 10-MHz laser-phase-modulation signal (see Fig. \ref{fig1} (c) for a typical resonance transmission spectrum and the corresponding PDH error signal). To calibrate the effectiveness of the electrodes, we define the electrode efficiency as $\eta = \frac{E_{\mathrm{mode}}}{U_0/d}$, where $E_{\mathrm{mode}}$ is the E-field strength along the Z direction at the location of the optical modes (see Appendix A for the simulated profiles of typical optical modes), $U_0$ is the voltage difference applied to the electrodes, and $d$ is the microresonator thickness. As shown in Fig. \ref{fig1} (d), we use finite-element-method (FEM) simulation \cite{cryst11030298} to compute the electrode efficiency, deriving $\eta \approx 0.3$. Next, we apply rf signals of sinusoidal waveforms with a peak-to-peak amplitude of 1 V on the electrodes and use a frequency-swept narrow-linewidth external-cavity diode laser to observe the transmission profile of a transverse-magnetic (TM) mode resonance with an optical bandwidth of approximately 1 MHz. The frequency of the rf signal is varied from 0.5 to 5 MHz. Figure \ref{fig1} (e) displays the spectrogram of the resonance, showing that modulation sidebands are generated. By comparing the depth of the central resonance and the sidebands we compute the resonance frequency deviation caused by the electrooptic modulation \cite{weng2024low}. Figure \ref{fig1} (f) shows the derived electrooptic modulation coefficients. We also utilize a vector network analyzer (VNA) to measure the frequency-dependent profile of the electrooptic modulation using the side slope of a low-$Q$ resonance to translate resonance frequency deviation into amplitude fluctuation of the transmitted light \cite{weng2024low}. The result is presented in the inset of Fig. \ref{fig1} (f), showing quantitative agreement with the characterization based on the modulation sidebands. Using the unclamped electrooptic coefficient $\gamma_{13}$ = 8.31 pm/V \cite{cryst11030298}, we estimate the frequency deviation caused by the applied electrical field with $\Delta f / f = \Delta n / n$ and \( \Delta n = \frac{1}{2} n^3 
\gamma_{ij} E_{\mathrm{mode}} \). Taking into account the electrode efficiency $\eta$, this estimated coefficient is in good agreement with the measured results at signal frequencies below 3 MHz. Above 3 MHz the modulation coefficient increases, reaching the peak of 1.7 MHz/V around signal frequency of 3.8 MHz. Such an enhancement of the electrooptic modulation is induced by the piezoelectric resonances \cite{takeda2012electro,weng2024low} of the bulk microresonator (see Appendix B for the simulation of the piezoelectric resonances), and can be utilized to improve the E-field sensitivity and the electrometry resolution.

\subsection{Electrometry sensitivity and resolution}

%%%%%%%%%%%%%%%%%%%%%%%%%%%%%%%%%%%%%%%%%
\begin{figure*}[htb]
\centering
%\hspace{20mm}
\includegraphics[width=2\columnwidth]{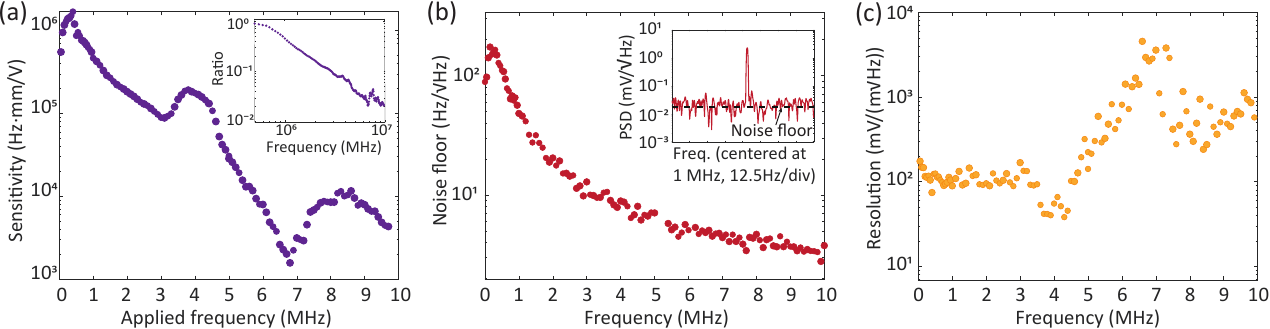} 
\caption{
(a) Electrometry sensitivity measured with the in-loop PDH error signal. The inset plots the ratio of the measured sensitivity to the electrooptic modulation coefficient displayed in Fig. \ref{fig1} (f), showing the cavity filtering effect.
(b) Noise floor of the PDH signal. The inset shows the power spectral density (PSD) of the voltage of the in-loop PDH error signal around signal frequency of 1 MHz.
(c) Electrometry resolution at varied signal frequencies.
}\label{fig2}
\end{figure*}
%%%%%%%%%%%%%%%%%%%%%%%%%%%%%%%%%%%%%%%%%

To activate the electrometry, a proportional-integral-derivative (PID) servo is employed to use the PDH error signal (with a slope of 0.50 V/MHz) to electrooptically lock the frequency of a microresonator resonance to the pump laser. The output voltage range of the servo is up to 100 V and the dc tuning coefficient of the resonance frequency is around 1 MHz/V, yielding a frequency control range of approximately 100 MHz. In our experiments the microresonator coupling setup is temperature stabilized so the PDH lock can be sustained indefinitely. In practical applications where temperature stabilization is not guaranteed, however, the drift of the resonance frequency may exceed the frequency control range, which would result in the loss of the PDH lock. The electrical field to be detected is generated with a sinusoidal rf signal from a function generator. The control signal from the servo is combined with the rf signal and applied to the microresonator electrodes using a bias-tee (see Fig. \ref{fig1} (a) and (b)). The peak-to-peak voltage of the rf signal is 10 mV and the signal frequency is varied from 0.1 to 9.95 MHz. The response to the rf signal from the in-loop PDH error signal is measured by an electrical spectrum analyzer (ESA) to analyze the frequency-dependent electrometry sensitivity (i.\,e., the frequency response to applied E-field strengths) and resolution (i.\,e., the detection limit related to the noise floor). As presented in Fig. \ref{fig2} (a), the sensitivity is computed by converting the measured amplitude of the error signal into frequency deviation using the calibrated PDH signal slope. Because of the optical cavity filtering effect, the sensitivity decreases as $\omega_{\text{s}}$ increases, with local rises and falls due to the complex interaction of piezoelectric resonances and intrinsic electrooptic effect (i.e., the piezoelectric resonances can either enhance or attenuate the electrooptic response depending on the relative phase). Figure \ref{fig2} (b) shows the measured noise floor at different signal frequencies. Again, the magnitude of the noise floor falls off with increased signal frequency. Using data presented in Fig. \ref{fig2} (a) and (b), the sensing resolution is computed and plotted in Fig. \ref{fig2} (c). At signal frequencies below 3 MHz, the resolution shows a flat profile that is not low-pass-filtered by the optical resonance, suggesting that the laser frequency noise is the dominant noise source according to our detailed theoretical analysis presented in Appendix C. Owing to the piezoelectric-resonance enhancement, at frequencies around 4 MHz the resolution is improved by nearly a factor of 3, reaching 34 $\text{mV}/(\text{m}\sqrt{\text{Hz}})$. Although such an enhancement is narrowband, in principle the frequencies of the piezoelectric resonances can be adjusted by modifying the geometry of the microresonator, which offers design flexibility if the targeted signal frequency falls within a relatively narrow range. Moreover, when used in practical antenna-free sensing, most of the attached electrodes should be removed (frequency locking with electrooptic control can be realized with smaller electrodes that cover only a small portion of the microresonator) so the sensitivity attenuation due to the low electrode efficiency is eliminated. As a result, the potentially achievable resolution is better than the results in Fig. \ref{fig2} (c) by a factor of $1/\eta = 3.3$.

\subsection{System noises and signal-to-noise ratio}

\begin{figure}[htb]
\centering
\includegraphics[width=0.96\columnwidth]{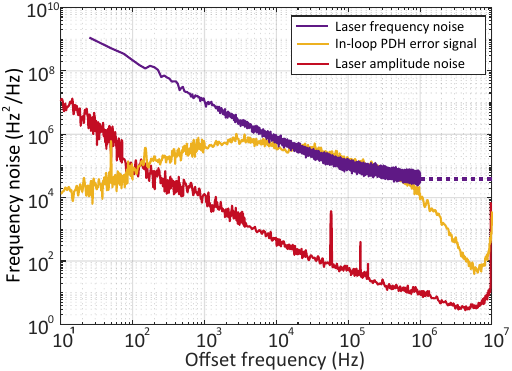} 
\caption{Noise spectra of system. Because the adopted self-heterodyne setup can only measure the laser noise at frequencies below offset frequency of 1 MHz, above this offset frequency we assume that the noise exhibits a white PSD profile as the flat dashed line. The spectra of the in-loop PDH signal and the laser amplitude noise show rises close to the offset frequency of 10 MHz, which is caused by the PDH phase modulation at 10 MHz.
}\label{fig3}
\end{figure}

To verify our deduction about the dominant noise floor, we measure the semiconductor laser frequency noise PSD using the short-delay self-heterodyne technique \cite{Zhao:22} and compare it with the noises of the PDH system that are measured at the output of the rf mixer. The noise spectra are plotted in Fig. \ref{fig3}. Within the PID servo control bandwidth of 3 kHz the in-loop error signal exhibits a much lower noise magnitude than that of the laser frequency noise. However, between 3 kHz and 1 MHz the two spectra overlap, indicating that the electrometry noise floor is dominated by the frequency noise of the pump laser. Above 1 MHz the noise magnitude of the error signal rapidly declines due to the optical cavity filtering. Assuming that the laser frequency noise spectrum is white above 1 MHz (see the dashed line in Fig. \ref{fig3}), after the application of the low-pass filtering due to the cavity filtering effect (see Appendix C for details), the laser frequency noise spectrum agrees very well with the measured in-loop PDH signal noise floor at signal frequencies beyond the optical resonance bandwidth, further confirming that the electrometry noise floor is laser-frequency-noise limited. We also measure the PDH noise spectrum caused by the laser amplitude noise by disengaging the microresonator from effective coupling. The result confirms that within the electrometry frequency range between 3 kHz and 10 MHz the laser amplitude noise is always well below the laser frequency noise.

%%%%%%%%%%%%%%%%%%%%%%%%%%%%%%%%%%%%%%%%%%%%
%%%%%%%%%%%%%%%%%%%%%%%%%%%%%%%%%%%%%%%%%

\begin{figure}[htb]
\centering
\includegraphics[width=0.98\columnwidth]{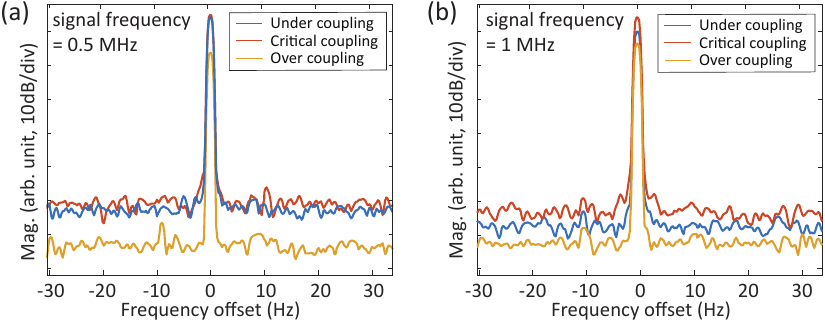} 
\caption{
ESA-measured signal spectra with different coupling conditions. The signal frequencies are 0.5 MHz for (a) and 1 MHz for (b), respectively.
}\label{fig4}
\end{figure}

Additionally, we test the electrometry signal-to-noise ratio (SNR) at certain signal frequencies with varied coupling loss. We change the gap between the prism and the microresonator so the coupling regime can be tuned from under-coupling to over-coupling, and we measure the SNR of rf signals of a constant amplitude of 10 mV using the ESA with a resolution bandwidth (RBW) of 500 Hz. Figure \ref{fig4} shows the rf spectra at two different signal frequencies of 0.5 and 1 MHz. Tuning the coupling regime changes the PDH signal slope due to the variations of the coupling efficiency as well as the loaded $Q$. As a result, the PDH sensitivities to both the microresonator resonance frequency deviations and the pump laser frequency noise are changed in the same manner. Both Fig. \ref{fig4} (a) and (b) show that the SNR stays the same despite that the signal's magnitude and the noise floor vary as the coupling regime is tuned. It reveals that the changing of the coupling regime has no impact to the electrometry resolution, which is expected from the laser-frequency-noise-dominated system. In other words, as long as the system is laser-frequency-noise limited, the performance cannot be improved by increasing the microresonator $Q$ (i.\,e., narrowing the optical resonance bandwidth).

%%%%%%%%%%%%%%%%%%%%%%%%%%%%%%%%%%%%%%%%%%%%
%%%%%%%%%%%%%%%%%%%%%%%%%%%%%%%%%%%%%%%%%

\subsection{Dynamic range}

\begin{figure}[hbt]
\centering
\includegraphics[width=0.98\columnwidth]{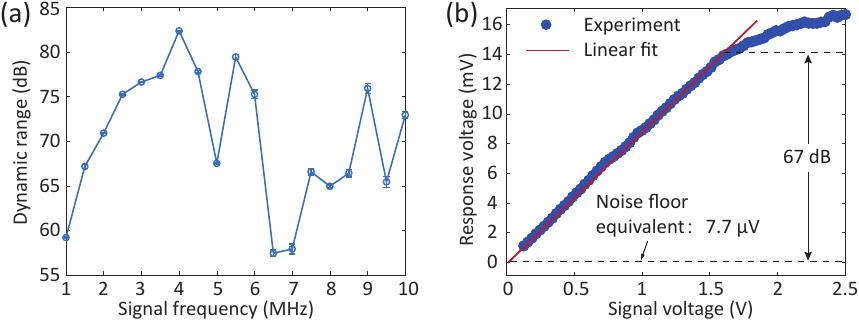} 
\caption{
(a) Dynamic range measured at discrete signal frequencies between 1 and 10 MHz. Error bars represent the standard deviation of 5 repeated measurements.
(b) The measured PDH signal response to an applied rf signal at 5 MHz with gradually increased amplitude. A dynamic range of 67 dB is obtained with this take.
}\label{fig5}
\end{figure}

We further calibrate the sensor's dynamic range at signal frequency range from 1 to 10 MHz. For each applied signal frequency, the amplitude of the rf signal is gradually increased. The PDH signal's response is monitored by the ESA to obtain the maximum signal amplitude when linearity is still maintained. With the measured noise floor we are able to compute the dynamic range. The result is displayed in Fig. \ref{fig5}. Within the measured frequency range the dynamic range reaches the maximum value of nearly 83 dB at 4 MHz and the minimum value of around 57 dB at 6.5 and 7 MHz, showing qualitative agreement with the separately measured sensitivity shown in Fig. \ref{fig2} (a).

%When the frequency exceeds 5 MHz, the system remains in the linear region within the tested voltage range(3.5V). Fig. \ref{fig6} (b) shows the detailed rf-amplitude-dependent response at the rf frequency of 5 MHz.Initially, the detected voltage increases linearly with the voltage applied to the electrode, eventually reaching the saturation voltage of 16 mV. The dynamic range of the linear regime is up to 67 dB. For the detected frequency, the signal noise floor is 7.7 $\mu$V (RBW = 1 Hz).

%When the E-field strength is too strong so the sidebands caused by the intracavity modulation exhibit deeper contrast that that of the central dip, the range of the PDH error signal is significantly compressed, setting the bottleneck of the dynamic range. This effect becomes more prominent when the rf frequency is higher than the optical resonance bandwidth because of the more resolved sidebands.

%%%%%%%%%%%%%%%%%%%%%%%%%%%%%%%%%%%%%%%%%%%%
%%%%%%%%%%%%%%%%%%%%%%%%%%%%%%%%%%%%%%%%%

%%%%%%%%%%%%%%%%%%%%%%%%%%%%%%%%%%%%%%%%%
%%%%%%%%%%%%%%%%%%%%%%%%%%%%%%%%%%%%%%%%%
%%%%%%%%%%%%%%%%%%%%%%%%%%%%%%%%%%%%%%%%%

\section{Conclusions}

In this work, we perform PDH-technique-based rf electrometry using a crystalline LN microresonator pumped by a low-cost semiconductor laser with a fixed lasing frequency and a relatively high frequency noise. Comprehensive characterization of the performance using a TM mode in the microresonator, including the sensitivity, resolution, SNR, and dynamic range, is carried out. Experiments with a transverse-electric (TE) mode are also performed, with results presented in Appendix D. The Pockels effect of the microresonator not only converts the external E-field strength into resonance frequency shift for low-noise signal readout but also allows for the effective frequency locking of the resonance to the laser. Leveraging the piezoelectric-resonance-enhanced electrooptic effect, our approach enables highly sensitive detection of E-field signals even the signal frequency is beyond the optical resonance bandwidth. The resolution achieved with this antenna-free scheme is already comparable to the performance of thin-film LN-waveguide-based interferometric sensors equipped with antenna arrays \cite{6188512,9189800}. This capability provides a significant step forward in overcoming the limitations of traditional optical resonator E-field sensors, offering wider applicability and enhanced versatility with a cost-effective setup. One should note that while the crystalline whispering-gallery-mode resonator used in this work may not be the most convenient microresonator type due to its relatively large size and the need for free-space optical coupling setup, the main system architecture can be easily transferred to integrated platforms \cite{Ma:2024aa,Yang_2024,Xia:25} to achieve a high level of robustness and a high spatial resolution (we refer readers to a comprehensive comparison of different E-field sensor platforms provided in \cite{Ma:2024aa} for detailed state-of-the-art electrometry performances). In addition, photonic sensors of other types, including thermometers and magnetometers, may also benefit from the developed scheme.

% \cite{boes2023lithium,Wang:2024aa}

%\newpage
\begin{acknowledgments}
This project is supported by Australian Research Council (DE210101904), ARC Centre of Excellence in Optical Microcombs for Breakthrough Science (COMBS), and University of Adelaide (DIGI+FAME Strategy Internal Grant 2021). Suwan Sun gratefully acknowledges financial support from China Scholarship Council.
\end{acknowledgments}

%%%%%%%%%%%%%%%%%%%%%%%%%%%%%%%%%%%%%%%%%%%%%%%%%%%%%%
%%%%%%%%%%%%%%%%%%%%%%%%%%%%%%%%%%%%%%%%%%%%%%%%%%%%%%
%%%%%%%%%%%%%%%%%%%%%%%%%%%%%%%%%%%%%%%%%%%%%%%%%%%%%%

\section*{Appendix A:  Simulation of optical modes}

\begin{figure}[h]
\centering
\includegraphics[width=1\columnwidth]{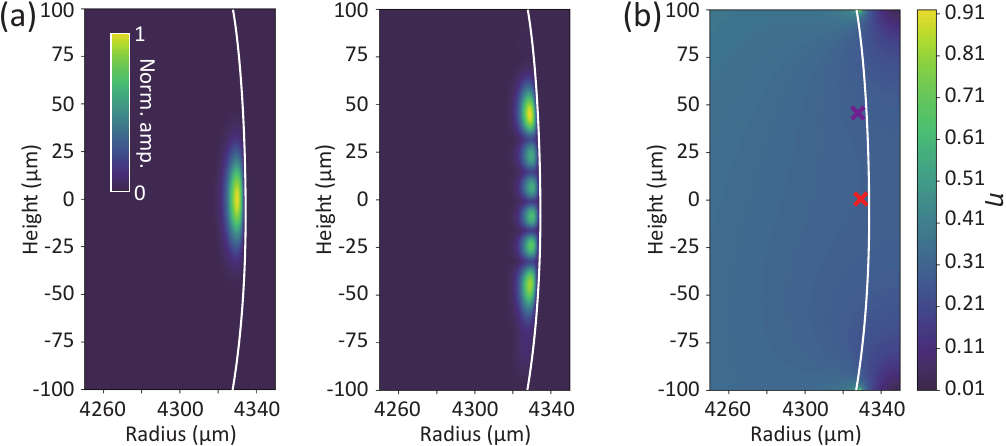} 
\caption{(a) Simulated electric field profiles of a fundamental TM mode (left) and a high-order TM mode (right).
(b) Simulated electrode efficiency around the locations of optical modes. The red and the purple crosses indicate the positions of the field intensity maxima of the two optical modes in (a), respectively. The electrode efficiencies at these two positions vary by approximately $3\%$.
}\label{fig6}
\end{figure}

We use FEM simulation to compute the cross-area optical mode profiles. Figure \ref{fig6} (a) presents two examples, including a fundamental TM mode and a high-order TM mode. The simulated electrode efficiency shows that these two modes have almost identical electrooptic sensitivities because the electrode efficiency varies by less than $4\%$ across the area that contains the most ($>99.5\%$) optical energy (see Fig. \ref{fig6} (b)). This is corroborated by our experimental tests with optical modes of different orders. Electrooptic sensitivities are measured with several different modes within a frequency range of the microresonator free spectral range. No notable difference in electrooptic sensitivity is observed.

 %%%%%%%%%%%%%%%%%%%%%%%%%%%%%%%%%%%%%%%%%%%%%%%%%%%%%%%%%%%%
%%%%%%%%%%%%%%%%%%%%%%%%%%%%%%%%%%%%%%%%%%%%%%%%%%%%%%%%%%%%

%%%%%%%%%%%%%%%%%%%%%%%%%%%%%%%%%%%%%%%%%%%%%%%%%%%%%%
%%%%%%%%%%%%%%%%%%%%%%%%%%%%%%%%%%%%%%%%%%%%%%%%%%%%%%
%%%%%%%%%%%%%%%%%%%%%%%%%%%%%%%%%%%%%%%%%%%%%%%%%%%%%%

\section*{Appendix B: Simulation of piezoelectric resonances}

\begin{figure}[h]
\centering
\includegraphics[width=1\columnwidth]{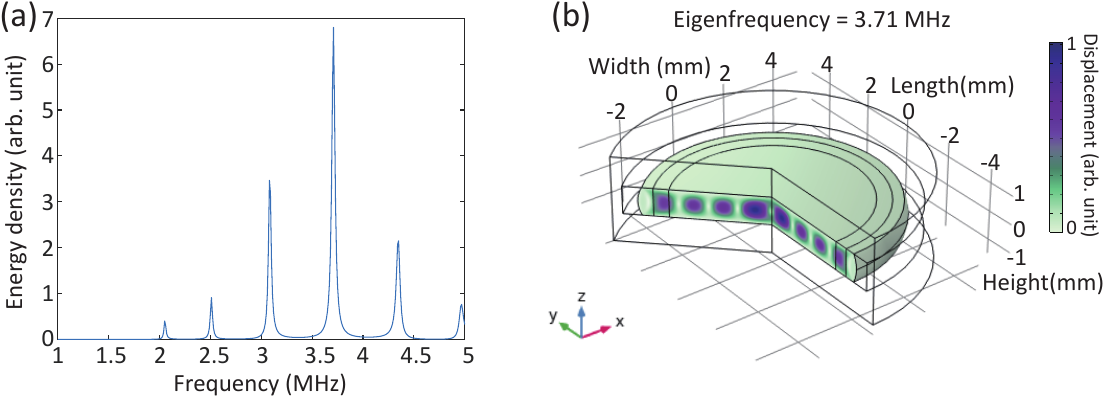} 
\caption{(a) Simulated piezoelectric resonances of the LN microresonator with FEM configuration of fixed boundaries and cylindrical symmetry. A low mechanical damping factor (0.01) is used to show narrow resonances. 
(b) The mechanical displacement profile of the strongest piezoelectric resonance at 3.71 MHz in (a).
}\label{fig7}
\end{figure}

Piezoelectric resonances of the LN microresonator is simulated with FEM method using the similar configuration employed in \cite{Weng:24}. Multiple resonances at the frequencies of a few MHz are revealed (see Fig. \ref{fig7} (a)), with the strongest resonance around 3.7 MHz (see the corresponding mechanical mode profile displayed in Fig. \ref{fig7} (b)), which is in good agreement with the experimental observations. We note that the simulation configuration of fixed boundaries and cylindrical symmetry would exclude some piezoelectric resonances that may contribute to the complex sensitivity behavior shown in Fig. \ref{fig2} (a). For the millimeter-scale size of the resonator the number of the allowed piezoelectric resonances within the relevant frequency range is large, and we are not able to experimentally verify the exact mode profiles of the observed strong resonances.

 %%%%%%%%%%%%%%%%%%%%%%%%%%%%%%%%%%%%%%%%%%%%%%%%%%%%%%%%%%%%
%%%%%%%%%%%%%%%%%%%%%%%%%%%%%%%%%%%%%%%%%%%%%%%%%%%%%%%%%%%%

\section*{Appendix C: Theory}

We analyze the microresonator mode frequency sensitivity and the laser-frequency noise floor of the PDH system using a derivation process similar to that of \cite{Ma:2024aa}. Using the pump laser's optical frequency as the rotating-wave reference frame and assuming that the laser field contains small-signal phase noise at the offset frequency of $\omega_\mathrm{s}$, the laser field is written as
%%%%%%%%%%%%%%%%%%%%%%%%%%%%%%%%%%%%%%%%%%%%%%%%%%%%%%%%%%%%%%%%
\begin{equation}
\label{eq1}
E_1 = E_0 \left[1 + i \mathcal{J}_1(\beta_1) e^{i \omega_\mathrm{s} t} + i \mathcal{J}_1(\beta_1) e^{-i \omega_\mathrm{s} t} \right]
\end{equation}
%%%%%%%%%%%%%%%%%%%%%%%%%%%%%%%%%%%%%%%%%%%%%%%%%%%%%%%%%%%%%%%%
where $E_0$ is the field of the laser carrier, $\beta_1$ is the index of the Bessel function that is related to the magnitude of the phase noise. After being phase-modulated at $\omega_\mathrm{m}$ for PDH signal generation, the optical field becomes
%%%%%%%%%%%%%%%%%%%%%%%%%%%%%%%%%%%%%%%%%%%%%%%%%%%%%%%%%%%%%%%%
\begin{equation}
\label{eq2}
E_2 = E_1 \left[\mathcal{J}_0(\beta_2) + i \mathcal{J}_1(\beta_2) e^{i \omega_\mathrm{m} t} + i \mathcal{J}_1(\beta_2) e^{-i \omega_\mathrm{m} t} \right]
\end{equation}
%%%%%%%%%%%%%%%%%%%%%%%%%%%%%%%%%%%%%%%%%%%%%%%%%%%%%%%%%%%%%%%%
where $\beta_2$ is related to the PDH phase modulation depth. 

The laser is coupled into a microresonator resonance when the resonance frequency is modulated by an external electrical field whose rf frequency is also $\omega_\mathrm{s}$. The dynamics of the intracavity light field $E$ can be studied by 
%%%%%%%%%%%%%%%%%%%%%%%%%%%%%%%%%%%%%%%%%%%%%%%%%%%%%%%%%%%%%%%%
\begin{equation}
\label{eq3}
\frac{dE}{dt} = -\left[ i \left( \Delta\omega + \omega(t) \right) + \frac{\kappa}{2} \right] E + \sqrt{\kappa_\mathrm{ex}} E_{\mathrm{in}}
\end{equation}
%%%%%%%%%%%%%%%%%%%%%%%%%%%%%%%%%%%%%%%%%%%%%%%%%%%%%%%%%%%%%%%%
where $E_{\mathrm{in}}$ is the pump field, $\Delta\omega$ is the detune of the laser frequency from the unmodulated resonance frequency, the resonance frequency modulation is denoted by $\omega(t) = \Omega \sin(\omega_\mathrm{s} t)$, where $\Omega$ is the frequency deviation that is dependent on the E-field strength and the frequency responsivity of the resonance. The cavity loss rate is denoted by $\kappa = \kappa_0 + \kappa_\mathrm{ex}$, where $\kappa_0$ is the intrinsic loss rate, and $\kappa_\mathrm{ex}$ is the external coupling loss rate.

\begin{figure}[h]
\centering
\includegraphics[width=1\columnwidth]{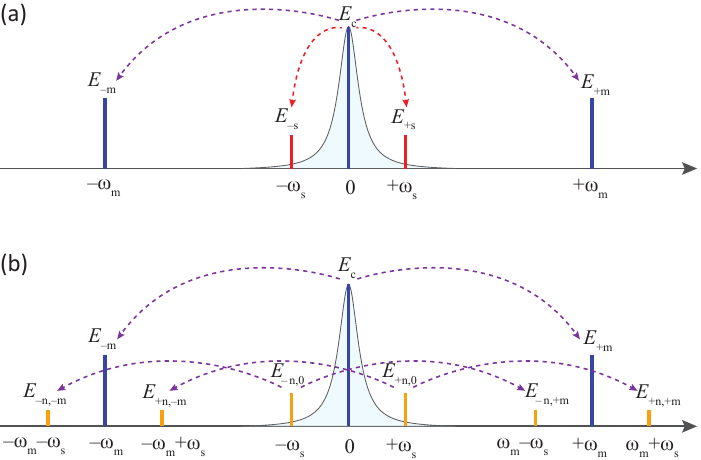} 
\caption{(a) Transmitted optical spectral components when a phase-modulated noiseless laser is coupled into the center of a frequency-modulated resonance. The dashed purple arrows represent the sideband generation process due to the PDH phase modulation. The dashed red arrows denote the sideband generation caused by the E-field-induced frequency modulation of the resonance.
(b) Transmitted optical spectral components when a laser with phase/frequency noise at the offset frequency of $\omega_\mathrm{s}$ is phase modulated at $\omega_\mathrm{m}$ and then coupled into a stationary resonance.
}\label{fig8}
\end{figure}

We first consider the response magnitude of the PDH signal due to the E-field-induced resonance frequency modulation. Because conventionally $\omega_\mathrm{m}$ is set to be much higher than the optical resonance bandwidth, only the central spectral component $E_0 \mathcal{J}_0 (\beta_2)$ in $E_2$ is coupled into the resonance when the PDH locking is engaged. With $\Delta\omega \approx 0$ due to the locking, the transmission of this spectral component $E_\mathrm{c}$ can be calculated using the steady-state solution of Eq.\ref{eq3} as
%%%%%%%%%%%%%%%%%%%%%%%%%%%%%%%%%%%%%%%%%%%%%%%%%%%%%%%%%%%%%%%%
\begin{equation}
\label{eq4}
E_\mathrm{c} = E_0 \mathcal{J}_0 (\beta_2) \left( 1 - \frac{2\kappa_\mathrm{ex}}{\kappa} \right)
\end{equation}
%%%%%%%%%%%%%%%%%%%%%%%%%%%%%%%%%%%%%%%%%%%%%%%%%%%%%%%%%%%%%%%%
In the meantime, as shown in Fig. \ref{fig8} (a), from the coupled-in light the E-field-induced resonance modulation produces two sidebands whose fields $E_{-\mathrm{s}}$ and $E_{+\mathrm{s}}$ in the transmission can be expressed as
%%%%%%%%%%%%%%%%%%%%%%%%%%%%%%%%%%%%%%%%%%%%%%%%%%%%%%%%%%%%%%%%
\begin{equation}
\label{eq5}
E_{-\mathrm{s}} = \frac{-i \kappa_\mathrm{ex} \Omega}{\kappa (\frac{\kappa}{2} - i \omega_\mathrm{s})} \mathcal{J}_0(\beta_2) E_0 e^{-i \omega_\mathrm{s} t}
\end{equation}
%%%%%%%%%%%%%%%%%%%%%%%%%%%%%%%%%%%%%%%%%%%%%%%%%%%%%%%%%%%%%%%%
%%%%%%%%%%%%%%%%%%%%%%%%%%%%%%%%%%%%%%%%%%%%%%%%%%%%%%%%%%%%%%%%
\begin{equation}
\label{eq6}
E_{+\mathrm{s}} = \frac{-i \kappa_\mathrm{ex} \Omega}{\kappa (\frac{\kappa}{2} + i \omega_\mathrm{s})} \mathcal{J}_0(\beta_2) E_0 e^{i \omega_\mathrm{s} t}
\end{equation}
%%%%%%%%%%%%%%%%%%%%%%%%%%%%%%%%%%%%%%%%%%%%%%%%%%%%%%%%%%%%%%%%

 %%%%%%%%%%%%%%%%%%%%%%%%%%%%%%%%%%%%%%%%%%%%%%%%%%%%%%%%%%%%
%%%%%%%%%%%%%%%%%%%%%%%%%%%%%%%%%%%%%%%%%%%%%%%%%%%%%%%%%%%%

With the two PDH phase-modulation sidebands that are expressed as:
%%%%%%%%%%%%%%%%%%%%%%%%%%%%%%%%%%%%%%%%%%%%
\begin{equation}
\label{eq7}
E_{-\mathrm{m}} = i \mathcal{J}_1(\beta_2) E_0 e^{-i \omega_\mathrm{m} t}
\end{equation}
%%%%%%%%%%%%%%%%%%%%%%%%%%%%%%%%%%%%%%%%%%%%
\begin{equation}
\label{eq8}
E_{+\mathrm{m}} = i \mathcal{J}_1(\beta_2) E_0 e^{i \omega_\mathrm{m} t}
\end{equation}
%%%%%%%%%%%%%%%%%%%%%%%%%%%%%%%%%%%%%%%%%%%%
we can calculate the optical power registered by the photodetector with the relevant terms that contain the modulation at $\omega_\mathrm{m} \pm \omega_\mathrm{s}$ as
%%%%%%%%%%%%%%%%%%%%%%%%%%%%%%%%%%%%%%%%%%%%%%%%%%%%%%%%%%%%
%%%%%%%%%%%%%%%%%%%%%%%%%%%%%%%%%%%%%%%%%%%%%%%%%%%%%%%%%%%%
\begin{equation}
\begin{split}
\label{eq9}
P_{\mathrm{s}}  =  & E_{+\mathrm{m}} E_{+\mathrm{s}}^* + E_{+\mathrm{m}} E_{-\mathrm{s}}^* + E_{-\mathrm{m}} E_{+\mathrm{s}}^* + E_{-\mathrm{m}} E_{-\mathrm{s}}^* \\
& + E_{+\mathrm{m}}^* E_{+\mathrm{s}} + E_{+\mathrm{m}}^* E_{-\mathrm{s}} + E_{-\mathrm{m}}^* E_{+\mathrm{s}} + E_{-\mathrm{m}}^* E_{-\mathrm{s}} \\
= & -\mathcal{J}_0 (\beta_2) \mathcal{J}_1 (\beta_2) P_0 \bigg[ 
\frac{2 \kappa_\mathrm{ex} \Omega}{\kappa \left(\frac{\kappa}{2} +  i \omega_\mathrm{s}\right)} e^{i (\omega_\mathrm{m} + \omega_\mathrm{s}) t} \\
& + \frac{2 \kappa_\mathrm{ex} \Omega}{\kappa \left(\frac{\kappa}{2} - i \omega_\mathrm{s}\right)} e^{i (\omega_\mathrm{m} - \omega_\mathrm{s}) t} + \frac{2 \kappa_\mathrm{ex} \Omega}{\kappa \left(\frac{\kappa}{2} + i \omega_\mathrm{s}\right)} e^{-i (\omega_\mathrm{m} - \omega_\mathrm{s}) t} \\
& + \frac{2 \kappa_\mathrm{ex} \Omega}{\kappa \left(\frac{\kappa}{2} - i \omega_\mathrm{s}\right)} e^{-i (\omega_\mathrm{m} + \omega_\mathrm{s}) t} \bigg] 
\end{split}
\end{equation}
%%%%%%%%%%%%%%%%%%%%%%%%%%%%%%%%%%%%%%%%%%%%%%%%%%%%%%%%%%%%
%%%%%%%%%%%%%%%%%%%%%%%%%%%%%%%%%%%%%%%%%%%%%%%%%%%%%%%%%%%%
where $P_0 = |E_0|^2$. This optical power, after optical-to-electrical conversion at the photodetection, is demodulated at $\omega_\mathrm{m}$ in a mixer to generate the PDH signal. The mixer's voltage output is
%%%%%%%%%%%%%%%%%%%%%%%%%%%%%%%%%%%%%%%%%%%%%%%%%%%%%%%%%%%%
%%%%%%%%%%%%%%%%%%%%%%%%%%%%%%%%%%%%%%%%%%%%%%%%%%%%%%%%%%%%
\begin{equation}
\label{eq10}
V_{\mathrm{s}} = P_{\mathrm{s}} \times a \cos(\omega_\mathrm{m} t)
\end{equation}
%%%%%%%%%%%%%%%%%%%%%%%%%%%%%%%%%%%%%%%%%%%%%%%%%%%%%%%%%%%%
%%%%%%%%%%%%%%%%%%%%%%%%%%%%%%%%%%%%%%%%%%%%%%%%%%%%%%%%%%%%
where $a$ is the optical-power-to-voltage conversion factor. Using $\cos(\omega_\mathrm{m} t) = \frac{e^{i \omega_\mathrm{m} t}}{2} + \frac{e^{-i \omega_\mathrm{m} t}}{2}$, and keeping only the terms with $e^{\pm i \omega_\mathrm{s} t}$, it becomes
%%%%%%%%%%%%%%%%%%%%%%%%%%%%%%%%%%%%%%%%%%%%%%%%%%%%%%%%%%%%
%%%%%%%%%%%%%%%%%%%%%%%%%%%%%%%%%%%%%%%%%%%%%%%%%%%%%%%%%%%%
\begin{multline}
\label{eq11}
V_{\mathrm{s}} = \frac{- 4 \kappa_\mathrm{ex} \Omega a \mathcal{J}_0 (\beta_2) \mathcal{J}_1 (\beta_2) P_0}{\kappa} \\
\left[ 
\frac{\frac{\kappa}{2}}{\left(\frac{\kappa}{2}\right)^2 + \omega_\mathrm{s}^2} \cos(\omega_\mathrm{s} t) 
+ \frac{\omega_\mathrm{s}}{\left(\frac{\kappa}{2}\right)^2 + \omega_\mathrm{s}^2} \sin(\omega_\mathrm{s} t) 
\right]
\end{multline}
%%%%%%%%%%%%%%%%%%%%%%%%%%%%%%%%%%%%%%%%%%%%%%%%%%%%%%%%%%%%
%%%%%%%%%%%%%%%%%%%%%%%%%%%%%%%%%%%%%%%%%%%%%%%%%%%%%%%%%%%%
As such, the detected signal electrical power is proportional to
%%%%%%%%%%%%%%%%%%%%%%%%%%%%%%%%%%%%%%%%%%%%
\begin{equation}
\label{eq12}
\left<|V_{\mathrm{s}}|^2\right> = \frac{32 \kappa_\mathrm{ex}^2 \Omega^2 a^2 \mathcal{J}_0^2 (\beta_2) \mathcal{J}_1^2 (\beta_2) P_0^2}{\kappa^2 (\kappa^2 + 4\omega_\mathrm{s}^2)}
\end{equation}
%%%%%%%%%%%%%%%%%%%%%%%%%%%%%%%%%%%%%%%%%%%%

Next, we derive the magnitude of the PDH signal response to the laser frequency/phase noise at the offset frequency $\omega_\mathrm{s}$ as in Eq. \ref{eq1}. As illustrated in Fig. \ref{fig8} (b), this laser frequency noise is manifested by two phase modulation sidebands in the frequency domain that are coupled into the microresonator resonance with frequency detunes of $\omega_\mathrm{s}$ and $-\omega_\mathrm{s}$, respectively. After the PDH phase modulation and the cavity filtering, the fields of the two sidebands in the transmitted light are expressed as:
%%%%%%%%%%%%%%%%%%%%%%%%%%%%%%%%%%%%%%%%%%%%
\begin{equation}
\label{eq13}
E_{-\mathrm{n},0} = i \left(1 - \frac{\kappa_\mathrm{ex}}{\frac{\kappa}{2} + i \omega_\mathrm{s}} \right) \mathcal{J}_1(\beta_1) \mathcal{J}_0(\beta_2) E_0 e^{-i \omega_\mathrm{s} t}
\end{equation}
%%%%%%%%%%%%%%%%%%%%%%%%%%%%%%%%%%%%%%%%%%%%
%%%%%%%%%%%%%%%%%%%%%%%%%%%%%%%%%%%%%%%%%%%%
\begin{equation}
\label{eq14}
E_{+\mathrm{n},0} = i \left(1 - \frac{\kappa_\mathrm{ex}}{\frac{\kappa}{2} - i \omega_\mathrm{s}} \right) \mathcal{J}_1(\beta_1) \mathcal{J}_0(\beta_2) E_0 e^{i \omega_\mathrm{s} t}
\end{equation}
%%%%%%%%%%%%%%%%%%%%%%%%%%%%%%%%%%%%%%%%%%%%
The PDH phase modulation also generates four additional spectral components originating from the frequency noise. Their fields are expressed as
%%%%%%%%%%%%%%%%%%%%%%%%%%%%%%%%%%%%%%%%%%%%
\begin{equation}
\label{eq5}
E_{-\mathrm{n},-\mathrm{m}} = - \mathcal{J}_1(\beta_1) \mathcal{J}_1(\beta_2) E_0 e^{-i (\omega_\mathrm{m} + \omega_\mathrm{s}) t}
\end{equation}
%%%%%%%%%%%%%%%%%%%%%%%%%%%%%%%%%%%%%%%%%%%%
\begin{equation}
\label{eq16}
E_{+\mathrm{n},-\mathrm{m}} = - \mathcal{J}_1(\beta_1) \mathcal{J}_1(\beta_2) E_0 e^{-i (\omega_\mathrm{m} - \omega_\mathrm{s}) t}
\end{equation}
%%%%%%%%%%%%%%%%%%%%%%%%%%%%%%%%%%%%%%%%%%%%
\begin{equation}
\label{eq17}
E_{-\mathrm{n},+\mathrm{m}} = - \mathcal{J}_1(\beta_1) \mathcal{J}_1(\beta_2) E_0 e^{i (\omega_\mathrm{m} - \omega_\mathrm{s}) t}
\end{equation}
%%%%%%%%%%%%%%%%%%%%%%%%%%%%%%%%%%%%%%%%%%%%
\begin{equation}
\label{eq18}
E_{+\mathrm{n},+\mathrm{m}} = - \mathcal{J}_1(\beta_1) \mathcal{J}_1(\beta_2) E_0 e^{i (\omega_\mathrm{m} + \omega_\mathrm{s}) t}
\end{equation}
%%%%%%%%%%%%%%%%%%%%%%%%%%%%%%%%%%%%%%%%%%%%%%%%%%%%%%
%%%%%%%%%%%%%%%%%%%%%%%%%%%%%%%%%%%%%%%%%%%%%%%%%%%%%%
%%%%%%%%%%%%%%%%%%%%%%%%%%%%%%%%%%%%%%%%%%%%%%%%%%%%%%
The optical power registered by the photodetector including the relevant terms with modulation at $\omega_\mathrm{m} \pm \omega_\mathrm{s}$ is written as
%%%%%%%%%%%%%%%%%%%%%%%%%%%%%%%%%%%%%%%%%%%%
%%%%%%%%%%%%%%%%%%%%%%%%%%%%%%%%%%%%%%%%%%%%%%%%%%%%%%
%%%%%%%%%%%%%%%%%%%%%%%%%%%%%%%%%%%%%%%%%%%%%%%%%%%%%%
\begin{equation}
\begin{split}
\label{eq19}
P_{\mathrm{n}}  = & E_\mathrm{c} E_{-\mathrm{n},-\mathrm{m}}^* + E_\mathrm{c} E_{+\mathrm{n},-\mathrm{m}}^* + E_\mathrm{c} E_{-\mathrm{n},+\mathrm{m}}^* + E_\mathrm{c} E_{+\mathrm{n},+\mathrm{m}}^*\\
& + E_{+\mathrm{m}} E_{-\mathrm{n},0}^* + E_{+\mathrm{m}} E_{+\mathrm{n},0}^* + E_{-\mathrm{m}} E_{-\mathrm{n},0}^* + E_{-\mathrm{m}} E_{+\mathrm{n},0}^*\\
& + E_{+\mathrm{n},0} E_{-\mathrm{m}}^* + E_{+\mathrm{n},0} E_{+\mathrm{m}}^* + E_{+\mathrm{n},+\mathrm{m}} E_{\mathrm{c}}^* + E_{+\mathrm{n},-\mathrm{m}} E_{\mathrm{c}}^* \\
& + E_{-\mathrm{n},0} E_{-\mathrm{m}}^* + E_{-\mathrm{n},0} E_{+\mathrm{m}}^* + E_{-\mathrm{n},+\mathrm{m}} E_{\mathrm{c}}^* + E_{-\mathrm{n},-\mathrm{m}} E_{\mathrm{c}}^*\\
= & 2 \kappa_\mathrm{ex} |E_0|^2 \mathcal{J}_1 (\beta_1) \mathcal{J}_1 (\beta_2) \mathcal{J}_0 (\beta_2)  [ (\frac{1}{\kappa/2} - \frac{1}{\kappa/2 - i \omega_\mathrm{s}}) e^{i (\omega_\mathrm{m} + \omega_\mathrm{s}) t} \\
& + (\frac{1}{\kappa/2} - \frac{1}{\kappa/2 + i \omega_\mathrm{s}}) e^{i (\omega_\mathrm{m} - \omega_\mathrm{s}) t} \\
& + (\frac{1}{\kappa/2} - \frac{1}{\kappa/2 + i \omega_\mathrm{s}}) e^{-i (\omega_\mathrm{m} + \omega_\mathrm{s}) t} \\
& + (\frac{1}{\kappa/2} - \frac{1}{\kappa/2 - i \omega_\mathrm{s}}) e^{-i (\omega_\mathrm{m} - \omega_\mathrm{s}) t} ]
\end{split}
\end{equation}
%%%%%%%%%%%%%%%%%%%%%%%%%%%%%%%%%%%%%%%%%%%%%%%%%%%%%%
%%%%%%%%%%%%%%%%%%%%%%%%%%%%%%%%%%%%%%%%%%%%%%%%%%%%%%
After demodulation, the PDH error signal voltage caused by the noise is
%%%%%%%%%%%%%%%%%%%%%%%%%%%%%%%%%%%%%%%%%%%%%%%%%%%%%%%%%%%%%%%%%%%%%%%%%%%%%%%%%%%%%%%%%%%%%%%%%%%%%%%%%%%%%%%%%%%%%%%%%%%%%%%%%%
%%%%%%%%%%%%%%%%%%%%%%%%%%%%%%%%%%%%%%%%%%%%%%%%%%%%%%
\begin{multline}
\label{eq20}
V_{\mathrm{n}} = 4 a \kappa_\mathrm{ex} P_0 \mathcal{J}_1 (\beta_1) \mathcal{J}_1 (\beta_2) \mathcal{J}_0 (\beta_2) \\
\left[ 
\cos(\omega_\mathrm{s} t) \left( \frac{1}{\kappa/2} - \frac{\kappa/2}{(\kappa/2)^2 + \omega_\mathrm{s}^2} \right) 
+ \sin(\omega_\mathrm{s} t) \left( \frac{\omega_\mathrm{s}}{(\kappa/2)^2 + \omega_\mathrm{s}^2} \right) 
\right]
\end{multline}
%%%%%%%%%%%%%%%%%%%%%%%%%%%%%%%%%%%%%%%%%%%%%%%%%%%%%%
%%%%%%%%%%%%%%%%%%%%%%%%%%%%%%%%%%%%%%%%%%%%%%%%%%%%%%
Assuming the laser frequency noise magnitude is offset-frequency-independent, we can use the substitution
%%%%%%%%%%%%%%%%%%%%%%%%%%%%%%%%%%%%%%%%%%%%%%%%%%%%%%
%%%%%%%%%%%%%%%%%%%%%%%%%%%%%%%%%%%%%%%%%%%%%%%%%%%%%%
\begin{equation}
\label{eq21}
\mathcal{J}_1 (\beta_1) \approx \frac{\beta_1}{2} = \frac{b}{2 \omega_\mathrm{s}}
\end{equation}
%%%%%%%%%%%%%%%%%%%%%%%%%%%%%%%%%%%%%%%%%%%%%%%%%%%%%%
%%%%%%%%%%%%%%%%%%%%%%%%%%%%%%%%%%%%%%%%%%%%%%%%%%%%%%
where $b$ is a constant frequency deviation value. As a result, the error signal voltage becomes
%%%%%%%%%%%%%%%%%%%%%%%%%%%%%%%%%%%%%%%%%%%%%%%%%%%%%%
%%%%%%%%%%%%%%%%%%%%%%%%%%%%%%%%%%%%%%%%%%%%%%%%%%%%%%
\begin{equation}
\label{eq22}
V_{\mathrm{n}} = \frac{2 a b \kappa_\mathrm{ex} P_0 \mathcal{J}_1 (\beta_2) \mathcal{J}_0 (\beta_2)}{\omega_\mathrm{s}} A \sin(\omega_\mathrm{s} t + \phi)
\end{equation}
%%%%%%%%%%%%%%%%%%%%%%%%%%%%%%%%%%%%%%%%%%%%%%%%%%%%%%
%%%%%%%%%%%%%%%%%%%%%%%%%%%%%%%%%%%%%%%%%%%%%%%%%%%%%%
with $\phi = \arctan(\frac{2\omega_\mathrm{s}}{\kappa})$ and $A = \frac{4}{\kappa} \sqrt{\frac{\omega_\mathrm{s}^2}{\kappa^2+4\omega_\mathrm{s}^2}}$. Eventually, we obtain that the signal power is proportional to
%%%%%%%%%%%%%%%%%%%%%%%%%%%%%%%%%%%%%%%%%%%%%%%%%%%%%%
%%%%%%%%%%%%%%%%%%%%%%%%%%%%%%%%%%%%%%%%%%%%%%%%%%%%%%
\begin{equation}
\label{eq23}
\left<|V_{\mathrm{n}}|^2\right> = 32 a^2 b^2 \kappa_\mathrm{ex}^2 P_0^2 \mathcal{J}_1^2 (\beta_2) \mathcal{J}_0^2 (\beta_2) \frac{1}{\kappa^2 (\kappa^2 + 4 \omega_\mathrm{s}^2)}
\end{equation}
%%%%%%%%%%%%%%%%%%%%%%%%%%%%%%%%%%%%%%%%%%%%%%%%%%%%%%
%%%%%%%%%%%%%%%%%%%%%%%%%%%%%%%%%%%%%%%%%%%%%%%%%%%%%%

Comparing Eqs. \ref{eq12} and \ref{eq23} shows that the PDH sensing signals caused by the rf electrical field and the laser frequency noise are attenuated by the microresonator resonance (i.e., cavity filtering effect) in the same manner. As a result, the electrometry SNR is not reduced for an rf electrical field whose frequency is beyond the optical resonance bandwidth, as long as the PDH system noise floor is dominated by the laser frequency noise. This result is intuitive because the PDH signal detects the relative frequency difference between the laser and the resonance.

%%%%%%%%%%%%%%%%%%%%%%%%%%%%%%%%%%%%%%%%%
%%%%%%%%%%%%%%%%%%%%%%%%%%%%%%%%%%%%%%%%%
\section*{Appendix D: Electrometry with a TE mode}

\begin{figure*}[htp]
\centering
\includegraphics[width= 1.5\columnwidth]{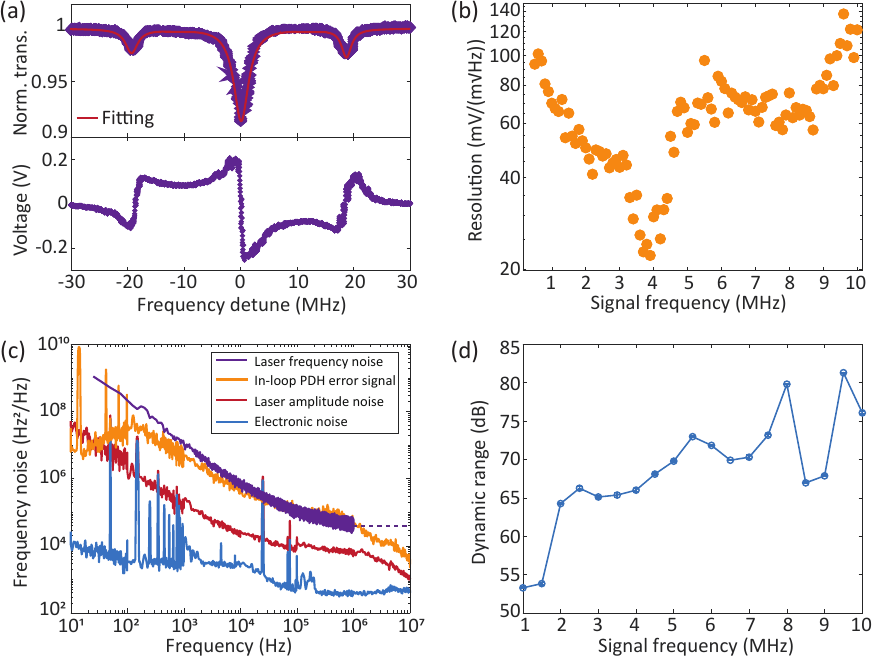} 
\caption{
(a) The transmission spectrum of a TE mode and the corresponding PDH signal.
(b) Measured electrometry resolutions with the TE mode.
(c) Laser frequency noise PSD measured with self-heterodyne technique and the PDH system noise spectra measured at the mixer output with the TE mode.
(d) Dynamic range at varied signal frequencies with the TE mode for electrooptic transduction.
}\label{fig9}
\end{figure*}

We change the coupling setup to perform the electrometry using a TE mode of the microresonator. Since the $Q$ of the mode resonance is lower ($\sim 5.1\times10^7$), we adopt a higher PDH modulation frequency of 20 MHz. The detune-frequency-swept spectroscopy of the resonance and its corresponding PDH signal are plotted in Fig. \ref{fig9} (a). Using the same procedures, the electrometry resolutions, the system noises, and the dynamic range are measured and presented in Fig. \ref{fig9} (b - d), respectively. Similarly, enhancement of the resolution due to piezoelectric resonances are observed, and the electrometry resolution is laser-frequency-noise limited. The achieved resolutions are in general better than those obtained with the TM mode because the involved electrooptic coefficient is larger (for the TE mode $\gamma_{33}$ = 28.6 pm/V). The best resolution is 22 mV/m$\sqrt{\mathrm{Hz}}$ at signal frequencies around 4 MHz, which means that the resolution in practical usage can reach below 7 mV/m$\sqrt{\mathrm{Hz}}$ when the rf attenuation due to the electrode is eliminated. We note that the overall dynamic range does not show improvement over those measured with the TM mode. With the larger optical resonance bandwidth, under intense electrooptic modulation the PDH error signal is severely distorted, resulting in a decreased frequency locking range and an unstable locking state. As such, the dynamic range is often limited by the loss of the PDH locking as the rf signal magnitude rises.

%%%%%%%%%%%%%%%%%%%%%%%%%%%%%%%%%%%%%%%%%%%%
%%%%%%%%%%%%%%%%%%%%%%%%%%%%%%%%%%%%%%%%%


%\clearpage
%\twocolumngrid
%\bibliography{TheReferences.bib}
\bibliography{Ref1}
%%%%%%%%%%%%%%%%%%%%%%%%%%%%%%%%%%%%%%%%%%%%
\end{document}